\documentclass[12pt]{iopart}
\usepackage{amssymb}
\usepackage{epsfig}
\usepackage{subfigure}
\usepackage{graphicx}
\usepackage{textcomp}
\usepackage{cite}
\usepackage{float}

\expandafter\let\csname equation*\endcsname\relax
\expandafter\let\csname endequation*\endcsname\relax
\usepackage{amsmath}
\newcommand{\agt}{\mathrel{\raise.3ex\hbox{$>$\kern-.75em\lower1ex\hbox{$\sim$}}}}

\begin{document}
\title{First Invader Dynamics in Diffusion-Controlled Absorption}

\author{S. Redner} \address{Department of Physics, Boston University, Boston,
  MA 02215, USA and Santa Fe Institute, 1399 Hyde Park Road, Santa Fe, New
  Mexico 87501, USA}

\author{Baruch Meerson} \address{Racah Institute of Physics, Hebrew
  University of Jerusalem, Jerusalem 91904, Israel}


\begin{abstract}

  We investigate the average time for the earliest particle to hit a
  spherical absorber when a homogeneous gas of freely diffusing particles
  with density $\rho$ and diffusivity $D$ is prepared in a deterministic
  state and is initially separated by a minimum distance $\ell$ from this
  absorber.  In the high-density limit, this first absorption time scales as
  $\frac{\ell^2}{D}\frac{1}{\ln\rho\ell}$ in one dimension; we also obtain
  the first absorption time in three dimensions.  In one dimension, we
  determine the probability that the $k^{\rm th}$-closest particle is the
  first one to hit the absorber.  At large $k$, this probability decays as
  $k^{1/3}\exp(-Ak^{2/3})$, with $A= 1.93299\ldots$ analytically calculable.
  As a corollary, the characteristic hitting time $T_k$ for the
  $k^{\rm th}$-closest particle scales as $k^{4/3}$; this corresponds to
  superdiffusive but still subballistic motion.
\end{abstract}

\maketitle

\section{Statement of the Problem}

\begin{figure}[ht]
\centerline{\includegraphics[width=0.35\textwidth]{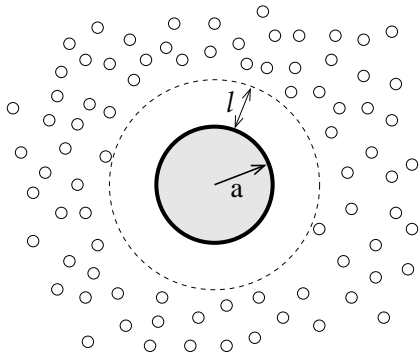}}
\caption{ Model geometry in two dimensions.  A gas of density $\rho$ occupies
  the region $r>a+\ell$ exterior to a circular absorber centered at the origin.}
\label{model-gen-d}
\end{figure}

Suppose that a gas of independent random walkers is uniformly and
deterministically distributed with density $\rho$ at time $t=0$ in the
exterior region that is a distance $\ell$ beyond a spherical absorber of
radius $a$ (Fig.~\ref{model-gen-d}).  For $t>0$, the particles diffuse freely
and are absorbed when they hit the absorber.  This flux, or reaction rate, is
fundamental to our understanding of many diffusion-controlled kinetic
processes (see e.g., \cite{S16,C43,B60,BP77,R85,OTB89}).  In this work, we
are interested in the behavior of this flux at short times.  Specifically:
(a) What is the average time for the earliest particle in the gas to first
hit the absorber?  (b) What is the probability that the $k^{\rm th}$-closest
particle hits the absorber first?

These first-hitting properties can be equivalently viewed as the survival
probability of a static absorber in the presence of a gas of density $\rho$
of diffusing particles with diffusivity $D$ that kill the absorber upon
reaching it.  This process, that is known as the ``scavenger'' or ``target''
problem has been extensively studied for an initial random (Poisson)
distribution of particles~\cite{ZKB83,T83,RK84,BZK84,BO87,BB03,FM12,BMS13}.
In one dimension, the survival probability asymptotically decays as
$\exp\big[-C_1 \rho (Dt)^{1/2}\big]$, where $C_1$ is a constant of order 1
that is exactly known.  In general spatial dimension $d$, the corresponding
survival probability is $\exp\big[-C_d\,\rho (Dt)^{d/2}\big]$ for $d<2$ and
$\exp(-C_d \,\rho a^{d-2} D t)$ for
$d>2$~\cite{ZKB83,T83,RK84,BZK84,BO87,BB03,FM12,BMS13}.  Various aspects of
the spatial distribution of the random walkers at the first absorption event
have been considered in Ref.~\cite{KMR10}.  Here we focus both on when the
absorber is first hit and on the complementary property of the identity of
the random walker that hits the absorber first.  In contrast to previous
works, we consider a \emph{deterministic} initial condition in which there is
a fixed number of particles at each lattice site.  We find that for $d=1$
this initial determinism leads to different asymptotic results for the
probability that the absorber has not yet been hit by time $t$ compared to
the previously-studied case of a random initial distribution of diffusing
particles.

In the next section, we derive the dependence of the average first hitting
time on system parameters in one and in three dimensions.  We then determine
the probability that the $k^{\rm th}$-closest particle is the first to hit
the absorber in one dimension.  We also show that this $k^{\rm th}$-closest
particle typically moves faster than diffusively, so that it actually can be
the particle that hits the absorber first, but slower than ballistically.

\section{First Hitting Time}

\subsection{One Dimension}

\begin{figure}[ht]
\centerline{\includegraphics[width=0.5\textwidth]{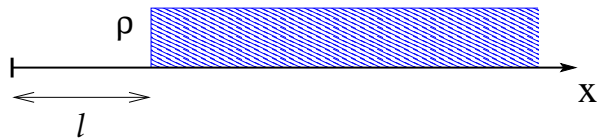}}
\caption{Model geometry in one dimension.  A gas of density $\rho$ occupies
  the region $x>\ell$, with an absorber at the origin.}
\label{model-1d}
\end{figure}

In one dimension, the radius of the absorber is irrelevant, and the resulting
system (Fig.~\ref{model-1d}) is characterized by three parameters: the
initial distance $\ell$ between the absorber and the closest particle, the
diffusion coefficient $D$, and the density $\rho$.  From these, we can form
two parameter combinations with units of time: $\ell^2/D$, the time to
diffuse from the closest particle to the absorber, and $1/(D\rho^2)$, the
time to diffuse between neighboring particles.  The first hitting time may
generally be written as
\begin{equation}
T = \frac{\ell^2}{D} \,F\left(\frac{\ell^2/D}{1/(D\rho^2)}\right) =
\frac{\ell^2}{D} \,f(\rho\ell)\,.
\end{equation}
Our goal is to determine the function $f$.

We begin by considering the gas to consist of a single particle at $x=\ell$.
The first-passage probability that this particle to first hit the origin at
time $t$ is~\cite{R01}
\begin{equation}
\label{f1}
f(\ell,t) = \frac{\ell}{\sqrt{4\pi Dt^3}}\, \,e^{-\ell^2/4Dt}\,.
\end{equation}
While the particle eventually hits the origin because $\int_0^\infty
f(\ell,t)\,dt=1$, the average time to reach the origin, $\int_0^\infty t\,
f(\ell,t)\,dt$, is infinite because of the $t^{-3/2}$ tail in the
first-passage probability.  From \eqref{f1}, the probability that the hitting
time for a single particle is longer than $t$ (equivalently the particle
survival probability up to time $t$), is
\begin{align}
  S(\ell,t) = \int_t^\infty f(\ell,t')\,dt' =
  \mathrm{erf}(\ell/\sqrt{4Dt})\,.
\end{align}
For a localized group of $\Delta n=\rho\,\Delta x$ particles that are
initially in the interval $[x,x+\Delta x]$, the probability that the hitting
time is longer that $t$ is $\big[S(x,t)\big]^{\Delta n}$.
Consequently, the cumulative probability that the hitting time exceeds $t$
for a gas that occupies the region $x\geq \ell$ is
\begin{align}
\label{Q}
Q(t)=\prod_{m=0}^\infty [S(\ell+m \Delta x,t)\big]^{\Delta n}&=
\exp\left\{ \Delta n\sum_{m=0}^\infty \ln S(\ell+m \Delta x,t)\right\} \nonumber \\
&\simeq \exp\left\{\rho\int_\ell^\infty\ln\big[\mathrm{erf}(x/\sqrt{4Dt})\big]\, dx\right\} \nonumber \\
&=\exp\left\{\rho\ell\sqrt{\tau}
  \int_{1/\sqrt{\tau}}^\infty\ln\big[\mathrm{erf}(z)\big]\, dz\right\}\equiv
\exp \left[\rho\,\ell\,\Phi(\tau)\right]\,.
\end{align}
Here we define the scaled time $\tau=4Dt/\ell^2$, the dimensionless variable
$z=x/\sqrt{4Dt}$, and
\begin{equation}
\label{phi}
\Phi(\tau)\equiv \sqrt{\tau}\int_{1/\sqrt{\tau}}^\infty\ln\big[
  \mathrm{erf}(z)\big] dz\,.
\end{equation}

The probability that a particle first hits the origin at time $t$ is
$q(t)=-\frac{\partial Q}{\partial t}$, so that the average first hitting time
is
\begin{equation}
  T =\int_0^\infty t\,\, q(t)\,dt =\int_0^\infty
  Q(t)\,dt = \frac{\ell^2}{4D}\int_0^\infty d\tau\,\exp\left[\rho\,\ell\,\Phi(\tau)\right]\,.
\end{equation}
The dependence of this first hitting time on system parameters may be
extracted from the asymptotic behaviors of $\Phi(\tau)$.  As shown in
\ref{app:Phi}, these are:
\begin{equation}
\label{phi-asymp}
\Phi(\tau)=
\begin{cases} -E_\infty\sqrt{\tau}+\ln \sqrt{\tau} +1-\ln \frac{2}{\sqrt{\pi}}
  +\ldots \,,&\qquad \tau\to\infty\,,\\
-\frac{\tau^{3/2}}{2\sqrt{\pi}}\, e^{-1/\tau}\,, &\qquad \tau\to 0\,,
\end{cases}
\end{equation}
with $E_\infty\equiv - \int_{0}^\infty
\ln\big[\mathrm{erf}(z)\big]\,dz=1.034415\ldots$ as computed numerically.
As a result, the controlling factor in the
cumulative distribution has the limiting behaviors:
\begin{equation}
    Q(\tau) \simeq
\begin{cases}
  \exp\big(-E_\infty\, \rho \ell\,\sqrt{\tau}\big)\,, &\qquad\qquad  \tau\gg 1\,,\\
   \exp\left(-\rho\,\ell\,\,\frac{\tau^{3/2}}{2\sqrt{\pi}}\,\,
    e^{-1/\tau}\right)\,, &\qquad \qquad \tau\ll 1\,.
\end{cases}
\end{equation}
It is worth emphasizing that the long-time behavior of $Q(\tau)$ is given by
$\exp\big(-C \rho \ell\,\sqrt{\tau}\big)$ in the case of \emph{randomly} (Poisson) distributed
particles at $t=0$, with $C={1}/{\sqrt{\pi}}=
0.564189\ldots$~\cite{ZKB83,T83,RK84,BZK84,BO87,BB03,FM12,BMS13}; our result is
qualitatively similar, except that the coefficient in the exponential is
$E_\infty= 1.034415\ldots$.  The distribution $Q(\tau)$ also has an
extraordinarily sharp double exponential cutoff for short times.  This makes
it extremely unlikely that the first particle hits the absorber much earlier
than the diffusion time, as reflected in the behavior of the average hitting
time given below.

Let us now focus on the average first hitting time in the interesting
high-density limit of $\rho\ell\gg 1$, for which
\begin{subequations}
\begin{equation}
\label{largerhol}
T \simeq \frac{\ell^2}{4D} \int_0^\infty d\tau\,
\exp\left(-\rho\,\ell\,\,\frac{\tau^{3/2}}{2\sqrt{\pi}}\,\,
  e^{-1/\tau}\right)\,,\qquad\qquad \rho\ell\gg 1\,.
\end{equation}
A non-vanishing contribution to the integral arises only where the argument
of the exponential is less than one.  Ignoring subdominant factors, this
condition is satisfied when $\tau< (\ln\rho\ell)^{-1}$.  In this regime, we
approximate the entire exponential factor by 1 and integrate over the region
$\tau< (\ln\rho\ell)^{-1}$ to give
\begin{equation}
T \sim \frac{\ell^2}{4D}\,\, \frac{1}{\ln\rho\ell}\,,\qquad\qquad \ln \,(\rho\ell)\gg 1\,.
\end{equation}
By accounting for the subdominant factor $\tau^{3/2}/(2\sqrt{\pi})$ in the
exponential in \eqref{largerhol}, a better estimate of the point where the
integrand is non-negligible is $\rho\,\ell\,\tau^{3/2}\,
e^{-1/\tau}/{2\sqrt{\pi}}=1$.  This gives the more accurate asymptotic
\begin{equation}
T =\frac{\ell^2}{4D \ln\left(\frac{\rho\ell}{2 \sqrt{\pi}}\right)}\,
\,\left[1+\frac{3\ln \ln \left(\frac{\rho \ell}{2 \sqrt{\pi}}\right)}
{2 \ln \left(\frac{\rho \ell}{2 \sqrt{\pi}}\right)} +\dots \right]\,, \qquad\qquad \rho\ell\gg 1\,.
\end{equation}
In this high-density limit, the average first hitting time $T$ is finite,
even though the average first hitting time for any individual particle is
infinite.  The first hitting time is of the order of the average time for the
closest particle to diffuse to the absorber that is modified by a logarithmic
factor.  Because of the logarithmic density dependence, a huge increase in
the density leads to only a modest decrease in $T$.  This weak dependence
shows that most particles, especially distant ones, play a negligible role in
the first hitting event, as expected intuitively.

In the opposite limit of $\rho\ell\ll 1$, we use the large-$\tau$ behavior of
$\Phi(\tau)$ in \eqref{phi-asymp} to give
\begin{equation}
T \simeq  \frac{\ell^2}{4D} \int_0^\infty d\tau\,
\exp\left(-\rho\,\ell\,E_\infty\sqrt{\tau}+\rho \ell \ln \sqrt{\tau} +
  \dots\right)\,,
\qquad\qquad \rho\ell\ll 1\,.
\end{equation}
Now the contribution to the integral is non-negligible when
$\rho\ell\sqrt{\tau}\sim 1$ or $\tau_*\sim (\rho\ell)^{-2}$.  This allows us
to neglect the logarithmic term in the exponent and leads to the estimate
\begin{equation}
\label{T-small-rl}
  T \simeq  \frac{1}{2 E_\infty^2 D \rho^2} =\frac{0.46728\ldots}{D
    \rho^2}~,
\qquad \qquad  \rho\ell\ll 1\,.
\end{equation}
\end{subequations}
This limit corresponds to the situation of no initial separation between the
gas and the absorber, namely, $\ell=0$.

\subsection{Three Dimensions}

In two dimensions and above, the radius $a$ of the absorber must be non-zero
so that the probability to hit it is non-zero.  In three dimensions, the
probability that a diffusing particle initially at radius $r>a$ eventually
hits the absorber is~\cite{BP77,R01}
\begin{equation}
\label{eventual3d}
h(r) = \frac{a}{r}~.
\end{equation}
While it is not certain that a single diffusing particle will eventually hit
the absorber, one of the particles from an infinite gas will.  To show this
fact, consider a gas of density $\rho$ that uniformly fills the exterior
space.  The probability that no particles with radii in the range $r$ and
$r+\Delta r$ will reach the absorber is
\begin{equation}
\Big(1-\frac{a}{r}\Big)^{4\pi\rho r^2 \Delta r}\,.\nonumber
\end{equation}
The probability $P$ that no gas particles reach the absorber is
\begin{equation}
P= \prod_{m=0}^\infty\Big(1-\frac{a}{a+m \Delta r}\Big)^{4\pi\rho r^2 \Delta r}.
\end{equation}
In the continuum limit
\begin{equation}
\ln P =\int_a^\infty 4\pi r^2 \ln\Big(1-\frac{a}{r}\Big)\,dr\,,\nonumber
\end{equation}
which diverges to $-\infty$ at the upper integration limit; thus it is
impossible for all particles to miss the absorber.  This fact is actually
obvious because this system evolves to a steady state with a constant average
diffusive flux to the absorber~\cite{S16,C43,B60,BP77,R85,OTB89}.

We now take advantage of two simplifications to reduce the three-dimensional
diffusion problem to one dimension.  First, the flux to the surface of a
spherical absorber can be calculated by replacing the true initial condition,
a delta-function at the initial particle location $(r_0,\theta_0,\phi_0)$, by
a normalized spherically-symmetric initial condition of radius $r_0$.  Then
we can use the classic device of reducing a spherically-symmetric
three-dimensional diffusion problem to one dimension \cite{C43,R01}.  For the
concentration $\rho(r,t)$ in three dimensions that depends only on the radial
coordinate, the quantity $u(r,t)=r\rho(r,t)$ obeys the one-dimensional
diffusion equation.  That is, if $\rho$ solves
\begin{subequations}
\begin{equation}
\label{c6:DE-3d}
\frac{\partial \rho}{\partial t}=D\nabla^2_{3d}\rho,
\end{equation}
where $\nabla^2_{3d}$ is the radial Laplacian operator in three dimensions,
then $u=r\rho$ satisfies
\begin{equation}
\label{c6:DE-1d}
\frac{\partial u}{\partial t}=D\nabla^2_{1d}u.
\end{equation}
\end{subequations}

For a single particle that is initially at $r=r_0$, the effective spherically-symmetric concentration can be
written as a sum of a Gaussian and an anti-Gaussian, with the latter accounting
for the absorbing boundary condition at $r=a$.  From this expression, the
first-passage probability to the surface of the absorber is given
by~\cite{C43,R01}
\begin{equation}
\label{f1-3d}
f(r_0,t) = \frac{a(r_0-a)}{r_0\sqrt{4\pi Dt^3}}\, \,e^{-(r_0-a)^2/4Dt}\,,
\end{equation}
and the corresponding survival probability up to time $t$ is
\begin{align}
  S(r_0,t) = 1-\frac{a}{r_0}\mathrm{erfc} \left(\frac{r_0-a}{\sqrt{4Dt}}\right)\,,
\end{align}
where $\mathrm{erfc}\,z=1-\mathrm{erf}\,z$.
Notice that $S(r_0, t\to \infty)=1-a/r_0$, consistent with Eq.~(\ref{eventual3d}).

For a gas of uniform density $\rho$ that lies beyond a minimal radius $\ell$,
with $\ell>a$, the number of particles within the range $r$ and $r+\Delta r$
is $\Delta n=4\rho \pi r^2 \Delta r$.  The expression for $Q(t)$ that is the
analog of Eq.~\eqref{Q} is
\begin{align}
\label{Q-3d}
Q(t)=\prod_{m=0}^\infty \big[S(\ell+m \Delta r,t)\big]^{\Delta n}&\simeq
\exp\left\{4\pi\rho\int_{a+\ell}^\infty r^2\ln\left[1-\frac{a}{r}\,\mathrm{erfc}\left(\frac{r-a}{\sqrt{4Dt}}\right)\right]
  dr\right\} \nonumber \\
&\equiv \exp \left[\rho\,\ell\,\Phi_{\rm 3d}(\tau)\right]\,,
\end{align}
where we now define
\begin{equation}
\label{phi-3d}
\Phi_{\rm 3d}(\tau)=
4\pi\,\sqrt{\tau}\int_{\frac{\ell}{(a+\ell)\sqrt{\tau}}}^\infty(\sqrt{\tau}\, \ell z+a)^2\ln\left(1-\frac{a}{\sqrt{\tau}\, \ell z\!+\!a}\,
\mathrm{erfc} \,z\right) dz\,,
\end{equation}
with $\tau=4Dt/\ell^2$ the scaled time and $z$ the dimensionless variable
$z=(r-a)/\sqrt{4Dt}$.  In the limit of $\tau\to\infty$ we can set the lower
integration limit to zero and neglect the term $a$ in the expressions
$\sqrt{\tau}\, \ell z\!+\!a$ in the integrand.  Since the argument of the
logarithm is very close to $1$, we obtain
\begin{equation}
\label{longt3d}
  \Phi_{\rm 3d}(\tau)\simeq -4\pi \tau a \ell \int_0^{\infty} z \,\mathrm{erfc} \,z \,dz \simeq -\pi \tau a \ell = -\frac{4 \pi a D t}{\ell}.
\end{equation}
Then Eq.~\eqref{Q-3d} yields, as expected, the long-time asymptotic
$Q(t)\simeq \exp(-4 \pi \rho a Dt)$, independent of $\ell$.

In the short-time $\tau \to 0$ limit, we can neglect the factors
$\sqrt{\tau} \ell z$ in the integrand, so that the integration reduces to
$$
\int_{A}^\infty \ln \,\mathrm{erf} \,z\,dz, \qquad \mbox{where}\quad  A=\frac{\ell}{(a+\ell) \sqrt{\tau}}\,.
$$
Because $\tau$ is small, $A\gg 1$, and we can use the large-$z$ expansion
$\ln \,\mathrm{erf} \,z \simeq -e^{-z^2}/(\sqrt{\pi} z)$.  Since the
integrand decays rapidly with $z$, we write $z=A+\epsilon$ so that
$e^{-z^2}\simeq e^{-A^2-2A\epsilon}$, put $z\simeq A$ in the denominator
of the integrand and then perform the integral in $\epsilon$.  The final
result for $\Phi_{\rm 3d}(\tau)$ as $\tau \to 0$ is
\begin{equation}
\label{shortt3d}
  \Phi_{\rm 3d}(\tau)\simeq -\frac{2\sqrt{\pi}  a^2 \tau^{3/2} (a+\ell)^2}{\ell^2}\,e^{-\frac{\ell^2}{(a+\ell)^2 \tau}}.
\end{equation}
Using this result in Eq.~\eqref{Q-3d} we obtain, in physical variables,
\begin{equation}\label{Q3dshorttime}
Q\simeq \exp\left[ -\frac{16 \sqrt{\pi} \rho \, a^2 (a+\ell)^2 (Dt)^{3/2}}{\ell^5}\,e^{-\frac{\ell^4}{4(a+\ell)^2 Dt}}\right]
\end{equation}

To compute $T$ in the regime when the gas density if large, and the
dimensionless time $\tau$ is small, we apply the same reasoning as in the
evaluation of the corresponding integral \eqref{largerhol} in one dimension
and now find, to lowest order
\begin{equation}
  T \sim \frac{\ell^4}{4D (a+\ell)^2}\,\frac{1}{\ln \frac{\rho a^2 \ell^2}{a+\ell}}\,,\qquad\qquad
 \ln \frac{\rho a^2 \ell^2}{a+\ell} \gg 1\,,\qquad\qquad\qquad d=3\,.
\end{equation}


In the opposite limit of very low density (corresponding to large $\tau$),
the average first hitting time is
\begin{equation}\label{lowdensity3d}
T\simeq \frac{\ell^2}{4D} \int_0^{\infty} \exp \left(-\pi \rho a \ell^2 \tau \right) \,d\tau =\frac{1}{4 \pi \rho a D}.
\end{equation}
The initial separation does not play a
role in this limit.

\section{Which Particle Hits First?}

\begin{figure}[ht]
\centerline{\includegraphics[width=0.45\textwidth]{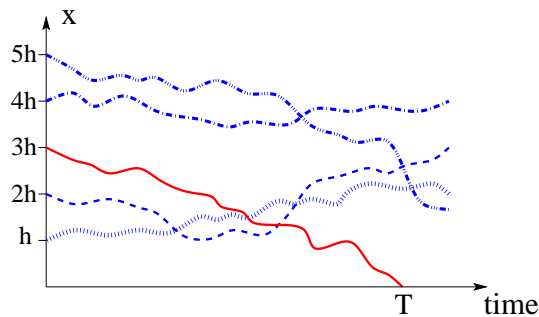}}
\caption{ Illustration of hitting in one dimension.  Particles are initially
  at $x=h,2h,3h,\ldots$.  The first particle (here particle 3) hits the
  absorber at time $T$.}
\label{model}
\end{figure}

We now determine the probability $G_k$ that the $k^{\rm th}$-closest particle
is the one that first hits the absorber (Fig.~\ref{model}).  We focus on the
one-dimensional system in which freely diffusing particles $1,2,3,\ldots$ are
at $x_1(t), x_2(t), x_3(t), \ldots$, with the initial condition of one
particle at each lattice cite: $x_k(t\!=\!0)=kh$, with $k\in\mathbb{I}$ and
$h$ is the lattice spacing.

As a preliminary, consider two particles, labeled 1 and 2, that are
initially at $x=h$ and $x=2h$.  The probability that particle 1 hits
the origin first may be written as
\begin{equation}
\label{G1}
G_1= \int_0^\infty dt\, f(h,t) \int_t^\infty dt'\, f(2h,t')=\int_0^\infty
dt\, f(h,t)\,\, \mathrm{erf}(2h/\sqrt{4Dt})\,,
\end{equation}
where $f(h,t)$ is the first-passage probability to the origin at time $t$
for a particle starting at $x=h$ (Eq.~\eqref{f1}).  Equation~\eqref{G1}
states that particle 2 must hit the origin after particle 1.  Defining
$z=h/\sqrt{4Dt}$ reduces the above integral to
\begin{equation}
\label{G1-sol}
G_1=\frac{2}{\sqrt{\pi}}\int_0^\infty e^{-z^2}\, \mathrm{erf}(2z)\, dz =
\frac{2}{\pi} \,\tan^{-1}2 = 0.704833\ldots
\end{equation}
If instead, the two particles are at $\ell$ and $\ell+h$, corresponding to
the finite gap $\ell$ between the absorber and the gas, then
$G_1=\frac{2}{\pi}\tan^{-1}\frac{\ell+h}{\ell}$, which approaches the
limiting value of $\frac{1}{2}$ as $\ell\to\infty$.

\begin{figure}[ht]
\centerline{\includegraphics[width=0.75\textwidth]{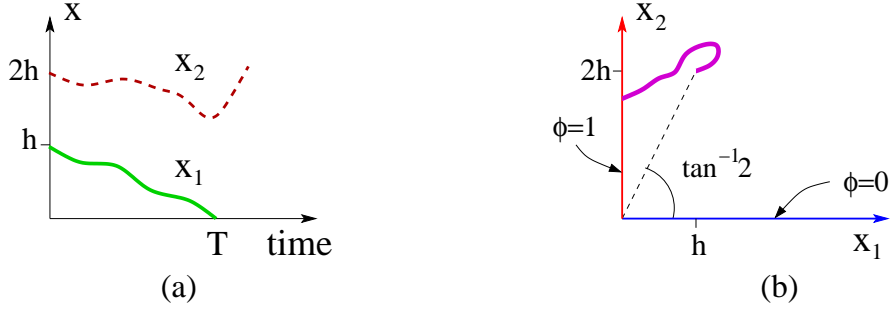}}
\caption{Correspondence between (a) two particles diffusing on the positive
  half line and (b) an effective particle, with coordinates $x_1$ and $x_2$,
  diffusing in the positive quadrant in two dimensions.}
\label{mapping}
\end{figure}

An alternative solution that involves minimal computation is to make a
correspondence (see Fig.~\ref{mapping}) between the two particles diffusing
on the half line to the diffusion of a single effective particle in two
dimensions that obeys suitable boundary conditions~\cite{FG88,R01}.  The
effective particle has initial coordinates $(x_1=h,x_2=2h)$ and is
constrained to remain within the positive quadrant $x_1,x_2>0$.  The
probability $G_1$ is equivalent to the effective particle first hitting the
line $x_1=0$, while $x_2$ always remains positive.  This hitting probability
equals the electrostatic potential at $(h,2h)$ when the line $x_1=0$ is held
at potential $\phi=1$ (corresponding to particle 1 first hitting the origin)
and the line $x_2=0$ is held at potential $\phi=0$ (corresponding to
particle 2 first hitting the origin)~\cite{BP77,R01}.  For this geometry, the
electrostatic potential at a point $(r,\theta)$ inside the quadrant is
$\phi(\theta)=\frac{2\theta}{\pi}$.  Because the point $(h,2h)$ corresponds
to polar angle $\tan^{-1}2$, the probability that particle 1 first hits the
boundary is just $\frac{2}{\pi}\tan^{-1}2$, as already given in
\eqref{G1-sol}.

To generalize to an infinite number of particles that are initially placed
deterministically at the lattice sites is conceptually straightforward.
Following the same steps that led to Eq.~\eqref{Q}, the analog of
Eq.~\eqref{G1} for an infinite gas is
\begin{align}
\label{G1-inf}
  G_1&= \int_0^\infty dt\, \frac{h}{\sqrt{4\pi Dt^3}}\,\, e^{-h^2/4Dt}
  \prod_{n=2}^\infty\,\, \mathrm{erf}(na/\sqrt{4Dt}) \nonumber \\
  &= \frac{1}{\sqrt{\pi}} \int_0^\infty \frac{d\tau}{\tau^{3/2}} \,\,
  e^{-1/\tau} \,\, \exp\Big\{ \sqrt{\tau}\int_{2/\sqrt{\tau}}^\infty dz \,
  \ln \big[\mathrm{erf}(z)\big]\Big\}\,.
\end{align}
We can extend \eqref{G1-inf} to the probability $G_k$ that the $k^{\rm th}$
particle hits the origin first:
\begin{align}
\label{G_k}
  G_k&= \int_0^\infty dt\, \frac{kh}{\sqrt{4\pi Dt^3}}\,\, e^{-(kh)^2/4Dt} \,\,
  \prod_{n=1}^{k-1}\,\, \mathrm{erf}(nh/\sqrt{4Dt})\,
  \times \prod_{n=k+1}^\infty\,\, \mathrm{erf}(nh/\sqrt{4Dt}) \nonumber \\
  &= \int_0^\infty dt\, \frac{kh}{\sqrt{4\pi Dt^3}}\,\,
\frac{e^{-(kh)^2/4Dt}}{\mathrm{erf}(kh/\sqrt{4Dt})} \,\,
  \Big[\prod_{n=1}^\infty\,\, \mathrm{erf}(nh/\sqrt{4Dt})\Big]\,.
\end{align}

\begin{figure}[ht]
\centerline{
\subfigure[]{\includegraphics[width=0.45\textwidth]{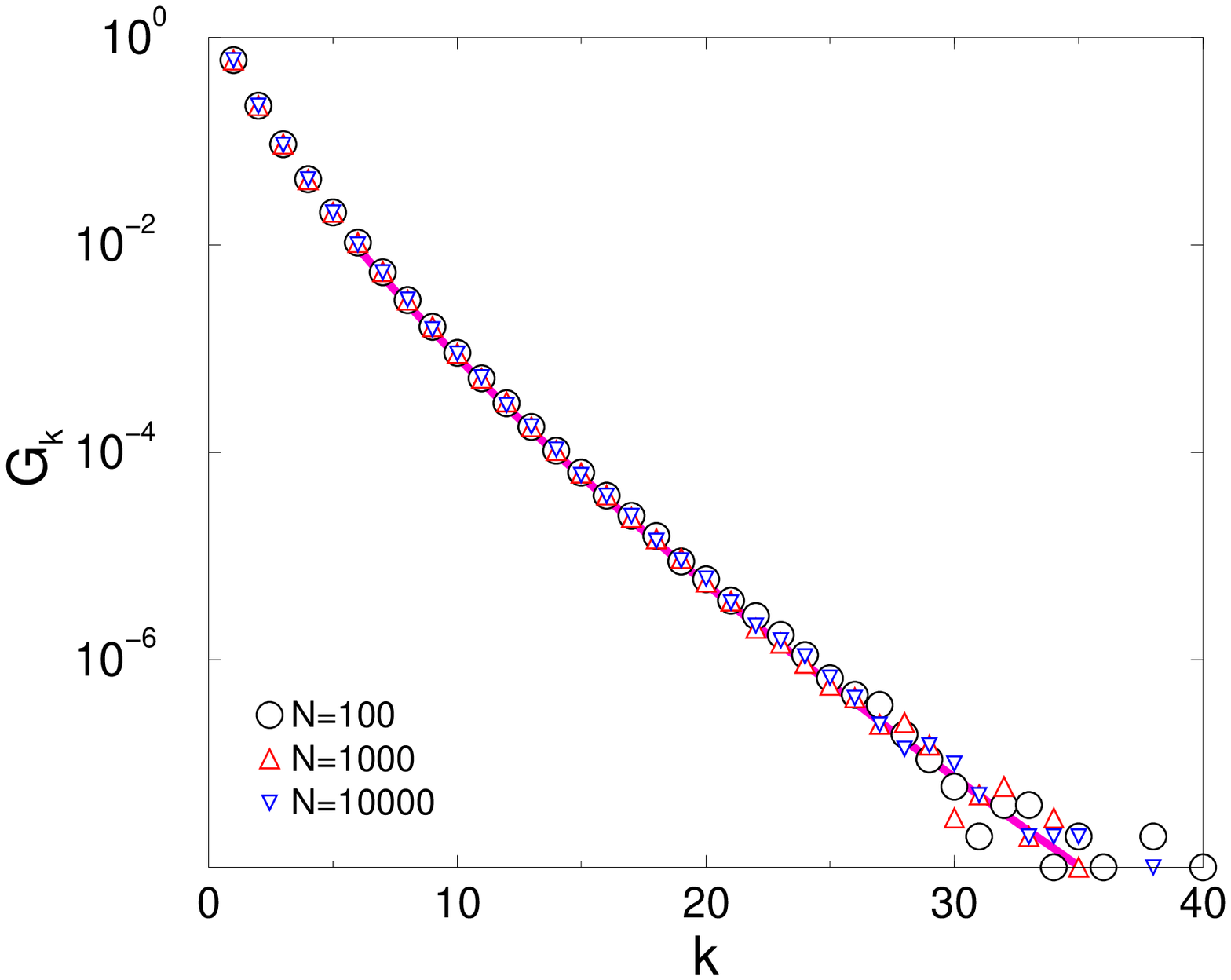}}\qquad
\subfigure[]{\includegraphics[width=0.43\textwidth]{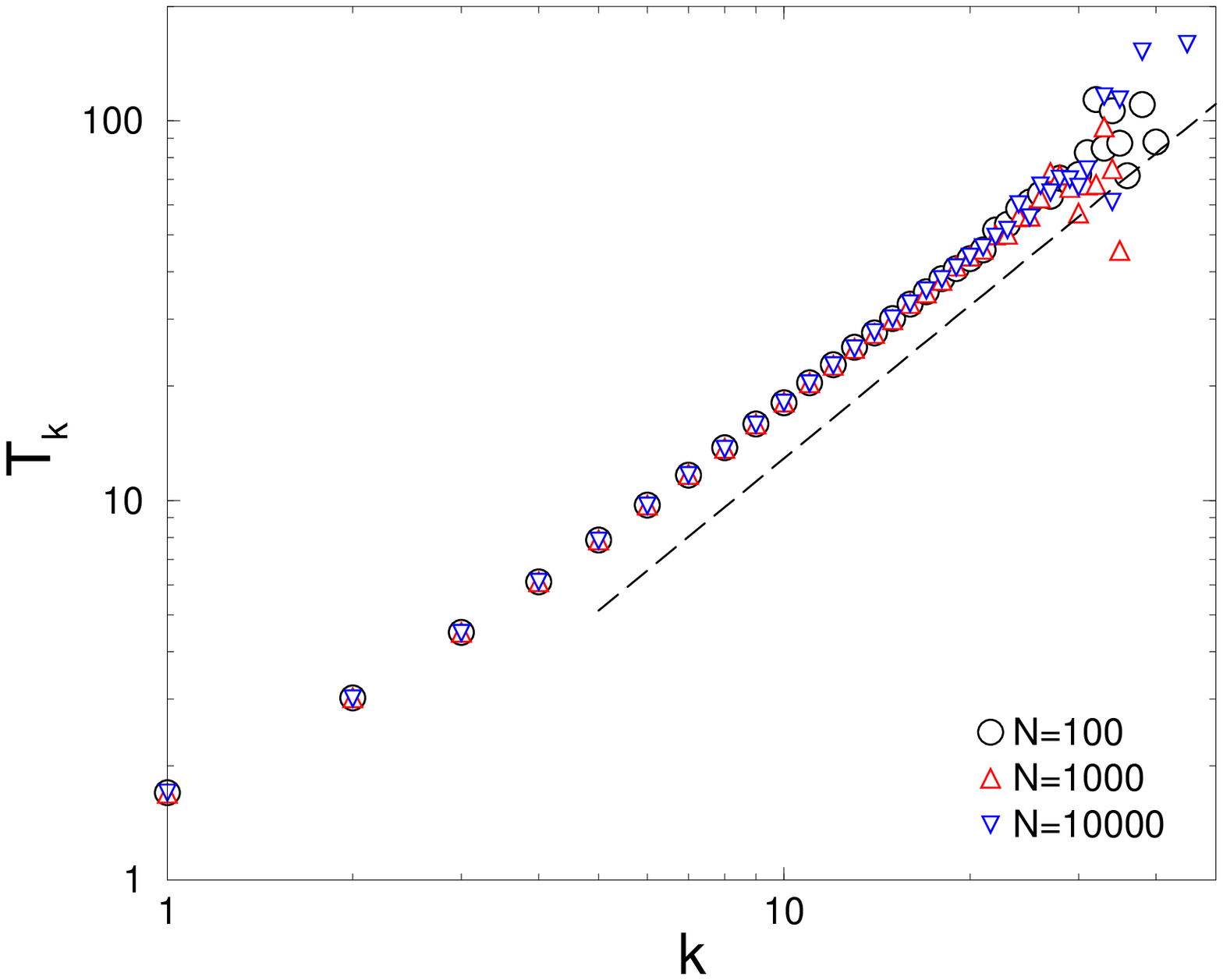}}}
\caption{(a) Probability that the $k^{\rm th}$ particle is the first one to
  hit the origin when initially $N$ particles are located at lattice
  sites for $N=10,10^2,10^3,10^4$.  The curve is our asymptotic estimate
  \eqref{Gk-asymp}.  (b) The average time for the $k^{\rm th}$-closest
  particle to hit the origin for those realizations where this $k^{\rm th}$
  particle is also the first one to hit the origin.  The dashed line
  corresponds to the power-law behavior $t_k\sim k^{4/3}$.  The data are
  based on $10^8$ realizations.}
\label{Gk-fig}
\end{figure}

We evaluate asymptotic behavior of this integral for $k \gg 1$ by first identifying the
asymptotic behavior of the integrand and then applying the Laplace method
(\ref{app:laplace}).  The result is:
\begin{equation}
\label{Gk-asymp}
    G_k \simeq \frac{2 \pi^{3/4}\, k^{1/3}}{\sqrt{3}\,
      (E_{\infty}/2)^{1/6}}\,\,\, e^{-3 (E_{\infty}/2)^{2/3}\,k^{2/3}}\,,\qquad\qquad k\gg 1.
\end{equation}
As shown in Fig.~\ref{Gk-fig}(a), this prediction is in excellent agreement
with numerical simulation data.  Surprisingly, this agreement arises even for
relatively small $k$.  In our simulations, the system contains $N$ particles,
with one particle at each lattice site $h,2h,\ldots Nh$.  In an
update event, one of the $N$ particles is chosen at random and moved by $\pm
h$, and the time is incremented by $\frac{1}{N}$.  This update is repeated
until one of the particles first hits the origin, where the identity of this
first invader and the hitting time are recorded.

A natural complement to the question of which particle first hits the origin,
is the characteristic time $T_k$ that is needed for the $k^{\rm th}$-closest
particle to hit the origin, given that this particle is the first one to hit.
From the Laplace evaluation of the integral in Eq.~\eqref{G_k}, the maximum
of the integral arises when $t\propto \frac{h^2}{D} k^{4/3}$.  This behavior
suggests 
that $T_k$ also scales as $k^{4/3}$.  While we are unable
to obtain appreciable data for $k\agt 40$ because of the extreme
improbability of a particle more distant than 40 being the first one to hit
the origin, our numerical results are consistent with $T_k\sim
\frac{h^2}{D}\, k^{4/3}$ (Fig.~\ref{Gk-fig}(b)).  This is faster than
diffusion, but slower than ballistic motion.  Faster than diffusive motion is
to be expected; if a distant particle is going to be the first invader, it
needs to do so quickly or else another particle that is initially closer to
the origin will be the first invader.  However, we do not have an intuitive
explanation for the anomalous $k$ dependence of this first invader time
$T_k\sim \frac{h^2}{D}\, k^{4/3}$.

\section{Concluding Remarks}

While the diffusive flux to an absorber is a classic and well-understood
quantity, we have uncovered new microscopic features of this flux.  We
focused specifically on the properties of the first particle to reach the
absorber.  For a gas that is initially separated from the absorber by a
distance $\ell$, the earliest hitting time is of the order of the diffusion
time, but modified by a logarithmic function of the gas density.
This weak dependence means that it requires an extremely high density of the
gas to reduce the first hitting time much below the diffusion time.

As is obvious, it is the closest particle that is the most likely to first
reach the absorber.  Nevertheless, the probability that a more distant
particle is the first one to reach the absorber is not negligible.  We
computed, in one dimension, the exact probability $G_k$ that the $k^{\rm
  th}$-closest particle to the absorber will be the one to first reach it.
Both analytically and from simulations, we found that the controlling factor
in this probability decays at large $k$ as the stretched exponential function
$G_k\sim \exp(-Ak^{2/3})$, with $A$ exactly calculable and whose numerical
value is $1.93299\ldots$.  Correspondingly, the characteristic time $T_k$ for
the $k^{\rm th}$ particle to hit the origin scales as $k^{4/3}$.  This is
much less than the diffusion time, which scales as $k^2$, but much larger
than the ballistic time, which scales as $k$.  As one might anticipate, a
distant particle must hit the origin quickly if it is going to be the first
one to reach the origin.

In our discussion of the hitting probability $G_k$ of the
$k^{\rm th}$-closest particle, we have assumed the deterministic initial
condition of a fixed number of particles at each lattice site.  If the
particles are initially placed with a Poisson distribution of separations,
$G_k$ still behaves asymptotically as $G_k\sim \exp(-Bk^{2/3})$, but with $B$
distinct from the above constant $A=1.93299\ldots\ldots$, and with a
different $k$-dependence in the pre-exponential factor~\cite{Paul}.

Finally, it is worth mentioning that the first hitting probability for an
$N$-particle system is, in principle, computable from a corresponding
electrostatic formulation.  The diffusion of the $N$ particles on the half
line is equivalent to the diffusion of a single effective particle in the
positive $2^{\rm N}$-tant in $N$-dimensional space~\cite{FG88,R01}.  The
probability that particle 1 first hits the origin equals the potential at the
initial point $(h,2h,3h,\ldots)$, with the plane $x_1=0$ held at potential
$\phi=1$ and all other planes $x_i=0$ held at zero potential.  While this
problem is simple to state, it does not seem to have a simple solution.

\medskip

We thank Paul Krapivsky for helpful discussions.  Financial support of this
research was provided in part by grant No.\ 2012145 from the United
States-Israel Binational Science Foundation (BSF) (SR and BM) and grant No.\
DMR-1205797 from the NSF (SR)

\appendix
\section{Asymptotics of $\Phi$}
\label{app:Phi}

When $\tau\gg 1$, we may set the lower integration limit to zero in
\eqref{phi} to give the leading long-time behavior $\Phi(\tau)\to
- E_\infty\sqrt{\tau}$, with $E_\infty=-\int_{0}^\infty
\ln\big[\mathrm{erf}(z)\big]\,dz=1.034415\ldots$.  At the next level of
approximation we define $\varepsilon \equiv 1/\sqrt{\tau}$ and write
\begin{equation}\label{sub1}
  \int_{\varepsilon}^\infty \ln\big[\mathrm{erf}(z)\big]\,dz =
\int_{0}^\infty \ln\big[\mathrm{erf}(z)\big]\,dz
-\int_0^{\varepsilon} \ln\big[\mathrm{erf}(z)\big]\,dz.
\end{equation}
For $z\ll 1$, we expand the error function in the last term as
$\mathrm{erf}(z) = 2z/\sqrt{\pi} + \dots$ so that
\begin{equation}
  \int_0^{\varepsilon} \ln\big[\mathrm{erf}(z)\big]\,dz = - \varepsilon
\Big[\ln \varepsilon-1+\ln\Big(\frac{2}{\sqrt{\pi}}\Big)+\dots \Big]\,,\nonumber
\end{equation}
which leads to the first line in Eq.~\eqref{phi-asymp}.

In the opposite limit of $\tau\ll 1$, we substitute the large-argument
expansion of the error function
\begin{equation}
\mathrm{erf}(z)= 1- \frac{e^{-z^2}}{\sqrt{\pi}\,z}+\ldots\,,\nonumber
\end{equation}
in the integral for $\Phi$ to give
\begin{equation}
  \Phi(\tau)\simeq -\sqrt{\tau}\int_{1/\sqrt{\tau}}^\infty \frac{1}{\sqrt{\pi}\, z}\,
  e^{-z^2}dz\,.\nonumber
\end{equation}
We estimate this integral by writing $z=\frac{1}{\sqrt{\tau}}+\epsilon$ and
expanding for small $\epsilon$ to give
\begin{equation}
  \Phi(\tau)\simeq  -\sqrt{\tau}\int_0^\infty \sqrt{\frac{\tau}{\pi}} \,
  \exp\left(-\frac{1}{\tau}-\frac{2\epsilon}{\sqrt{\tau}}\right)\,d\epsilon.
\end{equation}
Evaluating this integral leads to the second line in Eq.~\eqref{phi-asymp}.

\section{Asymptotic Estimate of $G_k$}
\label{app:laplace}

Starting with the second line of Eq.~\eqref{G_k} (in which the product has
been rewritten as the exponential of the sum)
\begin{align}
\label{exact1}
  G_k&= \int_0^\infty dt\, \frac{ka}{\sqrt{4\pi Dt^3}}\,\,
\frac{e^{-(kh)^2/4Dt}}{\mathrm{erf}(kh/\sqrt{4Dt})} \,\,
  \exp\left\{\sum_{n=1}^\infty\,\, \ln\left[\mathrm{erf}(nh/\sqrt{4Dt})\right]\right\}\,,
\end{align}
we introduce the  variable $u=h/\sqrt{4Dt}$, to rewrite Eq.~\eqref{exact1} as
\begin{equation}\label{exact2}
G_k=\frac{2 k}{\sqrt{\pi}}\,\int_0^{\infty} du\, \frac{e^{-(k u)^2}}{\text{erf} \,(k u)} \,\exp \Big\{\sum_{n=1}^{\infty} \ln \left[\text{erf}\,(n u)\right]\Big\}.
\end{equation}
For $k \gg 1$, the two dominant factors in the integrand are $e^{-k^2 u^2}$
and $e^{\Psi(u)}$, where $ \Psi (u)= \sum_{n=1}^{\infty} \ln
\left[\text{erf}\,(n u)\right]$.

The factor $e^{-(ku)^2}$ vanishes rapidly for $u\gg 1/k$, while $e^{\Psi}$
vanishes rapidly for $u\to 0$ so that we can evaluate the integral by the
Laplace method.  The determination of the asymptotic behavior of $\Psi(u)$ is
a bit involved because $\ln \text{erf}\,(nu)$ changes rapidly with $n$ for
small $u$.  Consequently, replacing the sum for $\Psi$ by an integral leads
to errors in the subleading terms that ultimately contribute to the power-law
prefactor in the expression for $G_k$.  Thus we split the sum as follows:
\begin{equation}
\label{Phiident}
    \Psi (u)= \sum_{n=1}^{N} \ln \left[\text{erf}\,(n u)\right]+
\sum_{n=N+1}^{\infty} \ln \left[\text{erf}\,(n u)\right]\,,
\end{equation}
and choose $N$ so that $N\gg 1$ but $N u\ll 1$. The latter inequality allows
us to replace $\text{erf}\,(nu)$ by $(2/\sqrt{\pi})\,nu$ in the first sum.
Then the first sum, which we denote by $\Psi_1(u,N)$, is
$$
\Psi_1 (u,N) = N \ln \frac{2}{\sqrt{\pi}}+N\ln u +\ln N !\,.
$$
Using the Stirling's formula for the last term leads to
\begin{equation}
\label{stirling}
\Psi_1 (u,N) \simeq N \ln \frac{2}{\sqrt{\pi}}+N\ln u +N \ln N-N+\ln \sqrt{2\pi N}\,.
\end{equation}
In the second sum in Eq.~(\ref{Phiident}), which we denote by $\Psi_2(u,N)$,
we can indeed replace the sum by the integral.  Accounting for the
Euler-Maclaurin correction, we obtain
\begin{equation}
  \Psi_2 (u,N) = \frac{1}{u} \int_{Nu}^{\infty} dz\, \ln (\text{erf}\,z)-
\frac{1}{2} \ln [\text{erf}\,(Nu)]
  \simeq \frac{1}{u} \int_{Nu}^{\infty} dz\, \ln
  (\text{erf}\,z)-\frac{1}{2} \ln\left(\frac{2Nu}{\sqrt{\pi}}\right).
  \label{EM}
\end{equation}

For the integral on the right hand side of Eq.~\eqref{EM} we can write
\begin{equation}
\int_{Nu}^{\infty} dz\, \ln (\text{erf}\,z) = \int_{0}^{\infty} dz\, \ln (\text{erf}\,z) -\int_{0}^{Nu} dz\, \ln (\text{erf}\,z) = -E_{\infty} -\int_{0}^{Nu} dz\, \ln (\text{erf}\,z).\nonumber
\end{equation}
The remaining integral can be easily evaluated at $Nu\ll 1$, by again using
$\text{erf}\,z = (2/\sqrt{\pi})\, z +\dots$, and we obtain
\begin{equation}\label{Phi2}
\Psi_2 (u) \simeq -\frac{E_{\infty}}{u}-N \ln(Nu)+N \ln\frac{\sqrt{\pi}}{2}+N
-\frac{1}{2} \ln\left(\frac{2Nu}{\sqrt{\pi}}\right).
\end{equation}
Adding the two contributions $\Psi_1$ and $\Psi_2$, we obtain
\begin{equation}
\Psi(u) \simeq -\frac{E_{\infty}}{u} -\ln \sqrt{u} +\ln\, \pi^{3/4}.\nonumber
\end{equation}
As it must, the result for $\Psi$ does not depend on the cutoff $N$.
Returning to Eq.~\eqref{exact2}, its large-$k$ asymptotic behavior is described by
\begin{equation}
\label{Gk-asymp-app}
G_k\simeq \frac{2 k}{\sqrt{\pi}}\,\int_0^{\infty} du\,\, \frac{e^{-(k
    u)^2}}{\text{erf} \,(k u)} \times \frac{\pi^{3/4}}{\sqrt{u}}\,\, e^{-E_{\infty}/u}\,.
\end{equation}
The product of the two competing exponents in the relevant region of $u$ can
be written as $e^{-H(u)}$, where
\begin{equation}\label{H}
    H(u)=(k u)^2 +\frac{E_{\infty}}{u}.
\end{equation}

The maximum contribution to the integral comes from the region near the
minimum of $H(u)$.  This occurs at $u_*=(E_{\infty}/2)^{1/3}\, k^{-2/3}$,
while the minimum value is $H(u_*)=3 (E_{\infty}/2)^{2/3}\, k^{2/3}$.  The
fact that $u_*\to 0$ as $k \to \infty$ justifies \emph{a posteriori} the
assumption $Nu\sim Nu_*\ll 1$ that we made in evaluating $\Psi(u)$.  Now we
expand
\begin{equation}\label{quadratic}
H(u) = H(u_*) +\tfrac{1}{2} H''(u_*)\,(u-u_*)^2 + \dots.
\end{equation}
Since $H''(u_*)=6 k^2$, the width of the peak of the integrand at $u=u_*$
scales as $\sim 1/k$, and so it is smaller (by a factor of $k^{1/3}\gg 1$)
than the peak position $u_* \sim k^{-2/3}$.  Thus the relative width of the
peak vanishes for $k\to\infty$.  Thus for large $k$ we can neglect
higher-order corrections in the expansion \eqref{quadratic}, expand the
integration range in \eqref{Gk-asymp-app} to $(-\infty,\infty)$, and evaluate
the slowly-varying factors $[\sqrt{u}\,\text{erf}\,(ku)]^{-1}$ at $u=u_*$.
Computing the resulting Gaussian integral gives final result that is quoted
in Eq.~\eqref{Gk-asymp}.  This formula is in excellent agreement with the
numerical evaluation of the exact expression (\ref{exact2}), with a relative
error that is less than 1\% already at $k=12$. As expected, there is
disagreement as small $k$. For example, for $k=1$ Eq.~\eqref{Gk-asymp} yields
$0.44011\ldots$, whereas the exact value that we computed numerically is
$G_1=0.6146 \ldots$.

\end{document}